\documentclass{PoS}

\title{Cosmic Rays Propagation with HelMod: Difference between \textit{forward-in-time} and \textit{backward-in-time} approaches}

\ShortTitle{Cosmic Rays Propagation with HelMod}

\author{ \speaker{S. Della Torre}$^a$, P. Bobik$^b$, M. J. Boschini$^{a,c}$, M. Gervasi$^{a,d}$, D. Grandi$^a$,  G. La Vacca$^a$,  S. Pensotti$^{a,d}$, M. Putis$^b$, P. G. Rancoita$^a$, D. Rozza$^{a,d}$,  M. Tacconi$^{a}$ and M. Zannoni$^{a,d}$\\
        \llap{$^a$}INFN Milano-Bicocca, \\
                   Piazza della Scienza,3 \\
                   20126 Milano - Italy\\
        \llap{$^b$}Department of Space Physics, Institute of Experimental Physics \\
                   Slovak Academy of Sciences\\
		   Watsonova 47, \\
		   040 01 Kosice, Slovakia\\
        \llap{$^c$}Cineca,\\
		   via R. Sanzio, 4 \\
		   20090 Segrate (MI) - Italy\\
        \llap{$^d$}Department of Physic, University of Milano-Bicocca \\
                   Piazza Ateneo Nuovo,3\\
                   20126 Milano - Italy\\
        E-mail: \email{stefano.dellatorre@mib.infn.it}}

\abstract{The cosmic rays modulation inside the heliosphere is well described by a transport equation introduced by Parker in 1965.
To solve this equation several approaches were followed in the past. 
Recently the Monte Carlo approach becomes widely used in force of his advantages with respect to other numerical methods. In the Monte Carlo approach, the transport equation is associated to a fully equivalent set of Stochastic Differential Equations. This set is used to describe the stochastic path of a quasi-particle from a source, e.g., the interstellar medium, to a specific target, e.g., a detector at Earth. In this work, we present both the Forward-in-Time and Backward-in-Time Monte Carlo solutions. We present an implementation of both algorithms in the framework of HelMod Code showing that the difference between the two approach is below 5\% that can be quoted as the systematic uncertain of the Method itself.
}

\FullConference{The 34th International Cosmic Ray Conference,\\
		30 July- 6 August, 2015\\
		The Hague, The Netherlands}
\newcommand{\vect}{\ensuremath{\vec}}
\newcommand{\matri}{\ensuremath{\tilde}}

\begin{document}
\section{Introduction}
Galactic Cosmic Rays (GCRs), entering into the heliosphere, experience of the so-called Solar Modulation: a reduction in the absolute flux, at energies $\lesssim 50$ GeV, with respect to the local interstellar spectrum (LIS). Particles propagating from the heliosphere boundary, located at $\sim 100$ AU, down to the Earth orbit have to pass through an expanding plasma emitted from the Sun (i.e. the Solar Wind). The small scale irregularities of the Sun magnetic field, carried out within the Solar Wind, causes a diffusion process of GCRs passing through the interplanetary medium.
The interplanetary conditions vary as a function of the solar cycle, that is approximately 22 years, consequently the intensity of the solar modulation is related to this cycle.
In general, particle propagation in the Heliosphere can be described using the well-known equation developed by Parker~\cite{parker1965}, based on a Fokker-Planck like Equation (FPE). 
The Parker's equation was initially solved using the so called ``Force Field'' approach~\cite{FFM1968,Gleeson1971,Gleeson1973}. In this quasi-analytical solution the whole diffusion/convection process can be described using a single parameter, i.e. the ``modulation potential''. 
Although this approximation is not able to reproduce all the physical processes in the heliosphere (see e.g. discussion in~\cite{Caballero2004}), due to its simplicity it remains the reference method for experimental treatment of Solar Modulation.
Numerical methods, e.g. Crank-Nicholson or finite difference method (see e.g.\cite{JokipiiKopriva1979,Potgieter85}), were used to solve the Parker's equation 
allowing to study in more detail the physics of the heliosphere.
The Heliosphere Modulation Model (HelMod)~\cite{Bobik2011ApJ,DellaTorre2013AdvAstro} implemented a new class of numerical methods, using a Monte Carlo technique. This is based on the mathematic equivalence between FPE and a set of Stochastic Differential Equation (SDEs) (see, e.g. Chapter 1.6-1.7 of~\cite{kloeden1999}). 
As many authors underline, see e.g.~\cite{JokipiiOwens1975,JokipiLev1977,Yamada1998,Zhang1999,alanko2007,PeiBurger2010,Bobik2011ApJ,Strauss2011}, this approach allows for more flexibility in model implementation, more stability of numerical results and to explore physical results that are \textit{hard to handle} with ``classical'' numerical methods or even not possible in the simple Force Field approach (see e.g.\cite{DellaTorre2012,DellaTorre2013AdvAstro,ICRC13_DellaTorre}).
In this work, we present two Monte Carlo solutions for the FPE applied on the problem of particle propagation in the heliosphere. The two solutions are obtained solving the FPE \textit{forward-in-time} and \textit{backward-in-time}. 

\section{Monte Carlo Method for HelMod}
The galactic cosmic rays transport equation was originally proposed by Parker in his fundamental work~\cite{parker1965,Jokipii1970,Fisk1971,Bobik2011ApJ}. 
This transport equation is a Fokker-Plank like equation that describes the modulation of Cosmic Rays by means of the so-called omni-directional distribution function $f(\vect{x},p)$ (see e.g.~\cite{JokipiiKopriva1979,Yamada1998,PeiBurger2010}) : 
\begin{equation}\label{eq::EQ_FPE}
 \frac{\partial f}{\partial t} = - \nabla \cdot (f\vect{V})
                                 + \nabla \cdot \left[ \matri K \cdot \nabla f\right]
                                 + \frac{(\nabla\cdot\vect V)}{3  p^2}\frac{\partial}{ \partial p}\left(p^3 f \right),
\end{equation}
where $p$ is particle momentum, $\vect{x}$ is the 3D-spatial position in Cartesian coordinates, $\vect{V} = \vect{V}_{sw} + \vect{V}_{drift}$, $\vect{V}_{sw}$ is the solar wind velocity, $\vect{V}_{drift}$ is the particle magnetic drift velocity and $\matri K$ is the diffusion tensor.
The differential intensity $J$ is related to $f$ as $J = p^{2} f$.
For sake of clarity, in this work, we treat the stochastic solutions of simple one dimensional Parker's equation in spherical coordinates. This allow us to focus directly on the numerical methods itself without adding to formulas the complexity of a more realistic treatment of the Heliosphere (see, e.g.,~\cite{DellaTorre2013AdvAstro} for a 2D \textit{forward-in-time} solution of Parker's equation using HelMod). In this approximation the diffusion tensor $\matri K$ was simplified to be a scalar $\kappa_{diff}$ and  all relevant description in the model (such as, e.g., the magnetic field and the solar wind) are spherically symmetric. Parker's equation can be thus simplified as follows: 
\begin{eqnarray}\label{eq::ParkerEQ_sph_1D_f}
 \frac{\partial f}{\partial t}&=&   \frac{1}{r^2}\frac{\partial  } {\partial r}\left(r^2 \kappa_{diff}\frac{\partial}{\partial r} f\right) - \frac{1}{r^2}\frac{\partial r^2 V_{sw} f}{\partial r} + \frac{1}{3}\left( \frac{1}{r^2}\frac{\partial r^2 V_{sw}}{\partial r} \right)\frac{1}{p^2}\frac{\partial}{\partial p}(p^3 f).
\end{eqnarray}
Since the magnetic field is assumed to be spherically symmetric, the magnetic drift velocity in radial direction is equal to zero. The Solar Wind ($V_{sw}$) is taken to be constant, radially directed and equal to $400$ km s$^{-1}$. The heliosphere in all approaches presented in this article is spherical with radius 100 AU and does not have Heliosheath or any other structure (Termination Shock, Heliopause, Bow Shock). In our Monte Carlo simulations we set an inner reflecting boundary at 0.3 AU.
  
With the Monte Carlo approach used in HelMod, the solution can be evaluated computing the SDE \textit{forward-in-time} and \textit{backward-in-time}. In \textit{forward-in-time} approach quasi-particles were traced in the heliosphere from the Heliosphere boundary down to the inner Heliosphere. In \textit{backward-in-time} approach the numerical process starts from the target, i.e. the Earth Orbit; quasi-particle objects are traced backward in time till the heliosphere boundary. In Ref. \cite{PeiBurger2010}, authors underline that a pseudo-particle is not a real particle nor a test particle; these are points in phase space that have the tendency to follow field lines according to SDEs, but they do not in general follow them rigorously owing to the random Wiener process present in the equation (see, e.g., ~\cite{PeiBurger2010}).

In term of differential operators, formally the \textit{backward-in-time} formulation is described by the adjoint operator of the \textit{forward-in-time}~\cite{Kopp2012}. This leads to the conclusion that, from the mathematically point of view, the two approaches describe the same process. 
However, in the knowledge of authors, in literature there is no contribution with direct quantitative comparison of spectra inside the heliosphere in order to have an estimation of a systematic error due to the numerical methods it self.

\subsection{Forward-in-time}
The \textit{forward-in-time} evolution of the transition density function $Q(s,y;t,x)$, from a phase-space point $y$ at time $s$ to a new position $(t,x)$, can be written as the follow FPE (see Equation 1.7.14 of~\cite{kloeden1999}, Equation 8 of~\cite{Zhang1999}, Equation 7 of~\cite{PeiBurger2010} and Equation 13 of~\cite{Kopp2012}):
\begin{equation}\label{eq:FPEGeneral_Fwd}
 \frac{\partial Q}{\partial t} = -\sum_i \frac{\partial}{\partial x_i}[A_{F,i}(t,x)Q]+\frac{1}{2}\sum_{i,j} \frac{\partial^2}{\partial x_i\partial x_j}[C_{F,ij}(t,x)Q] -L_F(t,x) Q + S 
\end{equation}
where $A_{F,i}$ is the advective term for \textit{i-th} coordinates (e.g. convection due to Solar Wind and Magnetic Drift), $C_{F,ij}$ is the diffusion tensor, $L_F$ is a bounded function that ``removes'' (or ``adds'') particles during the propagation (e.g. fragmentation, radioactive decay, see e.g.~\cite{Trottaetal2011}, Section 17.2.1 of~\cite{HandBookMC}) and finally $S$ represent a ``source'' of particles (e.g. Jovian Electrons, see~\cite{Strauss2011}). This equation gives the \textit{\textit{forward-in-time}} evolution with respect to the final state $(t,x)$.
In terms of SDEs, the evolution of the stochastic process can be described by (see e.g. Sections 4.3.2-4.3.5 of~\cite{gardiner1985} and Appendix A.13.1 of~\cite{HandBookMC}):
\begin{equation}\label{eq::SDEGeneral_Fwd}
 dx_i(t) = A_{F,i}dt+B_{F,i,j}dW_j(t),
\end{equation} 
where $\matri C_{F}=\matri B_{F}\matri B_{F}^T $, $d\vect W$ represents the increments of a \emph{standard Wiener process} that can be represented as an independent random variable of the form $\sqrt{d t} N(0,1)$ and, finally, $N(0,1)$ denotes a normally distributed random variable with zero mean value and unit variance~(See e.g., Appendix A of~\cite{Zhang1999} and Section 2 of \cite{Higham2001}). In term of stochastic propagation, the linear part $L_F$ is an additional parameter that allows the stochastic process to be created at an exponential rate as function of time~\cite{Zhang1999}. 

The Set of SDEs for the considered Parker's Equation can be obtained rewriting Eq.~\ref{eq::ParkerEQ_sph_1D_f} in the form of Eq.~\ref{eq:FPEGeneral_Fwd} (this can be done applying the substitution $F=r^2 f$ as indicated, e.g., in Refs.~\cite{Yamada1998,GervasiEtAl1999,PeiBurger2010,Bobik2011ApJ}). Then, using the equivalence with Eq.~\ref{eq::SDEGeneral_Fwd} the discrete \textit{forward-in-time} SDEs are the following (hereafter, we will refer to this as $\textbf{F-p}$ from \textbf{F}orward SDE with momentum $\textbf{p}$):
  \begin{eqnarray}
   \Delta r &=& \left(\frac{2 \kappa_{diff}}{r} + V_{sw}\right) \Delta t + N(0,1)\sqrt{(2\kappa_{diff} \Delta t)} \label{eq::F-pa}\\
   \Delta p &=& - \frac{2V_{sw} p }{3r}  \Delta t \label{eq::F-pb}\\
   L_F &=&- \frac{4 V_{sw}}{3 r} \label{eq::F-pc}
 \end{eqnarray}
 
In the Monte Carlo Forward-in-time approach the stochastic path described by Eqs.~\ref{eq::F-pa}-\ref{eq::F-pc} is followed by  quasi-particle objects injected at heliosphere boundary. The initial momentum distribution is chosen in order to sample the Local Interstellar Spectrum (LIS) distribution at heliosphere border. According to the sampling method, to each quasi-particle is associated a weight $w_0$ that reflect the LIS distribution probability. 
This weight is modified each $j$-th step during the propagation (from generation $j=0$ to registration $j=k$) in order to account for the Linear term $L$ of Eq.~\ref{eq:FPEGeneral_Fwd} (see Equation 22 in~\cite{Kopp2012}):
\begin{equation}\label{eq::pathweight}
W =w_0 \exp( -\sum_{j=0}^k L_{F,j}  \Delta t )
\end{equation}
When a quasi-particle pass through a selected position, momentum and weight, i.e., the one computed with Eq.~\ref{eq::pathweight}, are registered.
The modulated spectrum is obtained using a proper normalization scaling factor on the binned registered distribution. This is equivalent to apply the algorithm 17.2 of~\cite{HandBookMC}. 

\subsection{Backward-in-time}
The \textit{Backward-in-time} evolution of a FPE is described by the \textit{Kolmogorov backward equation} (see Equation 1.7.15 in Ref.~\cite{kloeden1999}). This equation gives the backward evolution of the associated transition density function with respect to the initial state $(s,y)$ (se e.g. Equation 13 of~\cite{Zhang1999}, Equation A2 of~\cite{Strauss2011}, Equation 14 of~\cite{Kopp2012}):
\begin{equation}\label{eq:FPEGeneral_Bck}
 \frac{\partial Q}{\partial s} = +\sum_i A_{B,i}(s,y) \frac{\partial Q}{\partial y_i}+\frac{1}{2}\sum_{i,j} C_{B,ij}(s,y) \frac{\partial^2 Q}{\partial y_i\partial y_j} -L_B Q + S 
\end{equation}
The meaning of the terms in Eq.~\ref{eq:FPEGeneral_Bck} are the same of Eq.~\ref{eq:FPEGeneral_Fwd}. These are here presented with different subscript since the diffusion and advective coefficients of forward and backward approach are the same only for the special case of constant value~\cite{Kopp2012}.
Thus, Eq.~\ref{eq::ParkerEQ_sph_1D_f} is associated with the following set of SDE for the \textit{backward-in-time} evolution (hereafter we will refer to this as $\textbf{B-p}$ from \textbf{B}ackward SDE with momentum $\textbf{p}$):
\begin{eqnarray}
   \Delta r &=& \left(\frac{2 \kappa_{diff}}{r} - V_{sw}\right) \Delta s + N(0,1)\sqrt{(2\kappa_{diff} \Delta s)} \label{eq::B-pa}\\
   \Delta p &=&  \frac{2V_{sw} p }{3r}  \Delta s \\
   L_B &=& 0 \label{eq::B-pc}
 \end{eqnarray}
We underline that $\Delta s>0$ represent a step \textit{backward-in-time}. 
One should note that Eqs.\ref{eq::B-pa}-\ref{eq::B-pc} differs from ~\ref{eq::F-pa}-\ref{eq::F-pc} for the follow: a) the convective term including solar wind reversed the sign, this is equal to a inward directed solar wind speed, b) in the $\textbf{B-p}$ case, quasi-particles gains momentum while propagating in the heliosphere and c) in F-p case $L_F>0$ means that, during the stochastic propagation, particles are exponentially ``removed'' due to the non-zero divergence of solar wind.

For each initial momentum $p_{detector}$, the modulated spectrum, $J_{modulated}$, is obtained as the average of un-modulated spectrum ($J_{LIS}$) evaluated at boundary reconstructed momentum ($p$)~\cite{Strauss2011,Kopp2012,HandBookMC}:
\begin{equation}
 J_{modulated}(p_{detector}) = \frac{p_{detector}^2}{N}\sum_{k=1}^N\left[ \frac{J_{LIS}(r_{bound},p)}{p^2}   \exp( -\sum_{j=0}^m L_{B,j}  \Delta s )\right]
\end{equation}
where $N$ is the number of generated quasi-particles with initial momentum $p_{detector}$, $m$ is the number of step occurred during the stochastic propagation of $k$-th quasi-particle and $r_{bound}$ is the exit point of stochastic path, i.e. the heliosphere boundary. 

\section{Numerical Solution}
\begin{figure}[htb]
\begin{center}
 \includegraphics[width=1.0\textwidth]{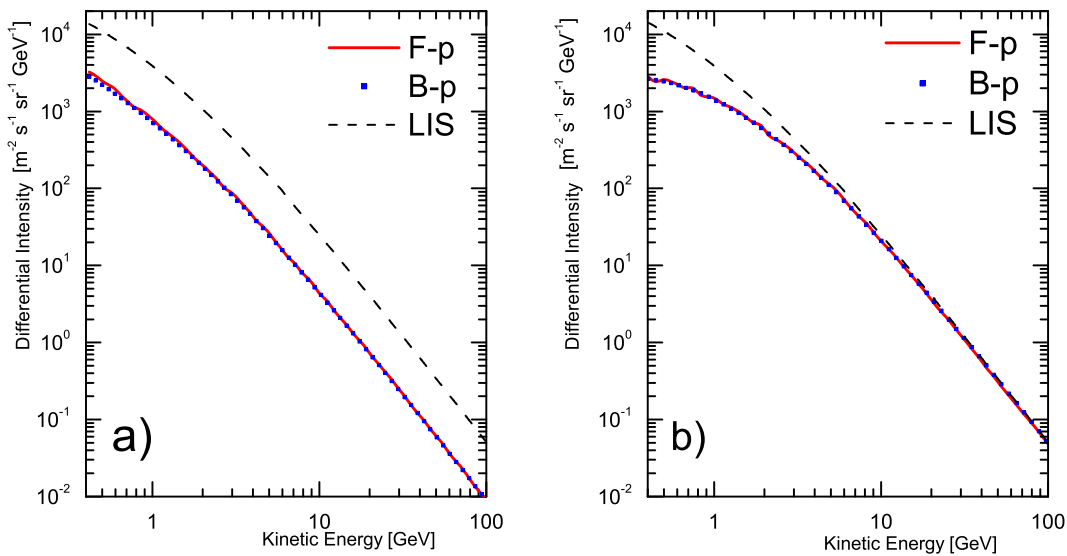}
 \caption{Monte Carlo solutions obtained with the F-p  (red line) and B-p (blue dot). In figure a) the diffusion coefficient is taken to be independent from rigidity (see eq.~(3.1)),
 i.e., the diffusion process is the same at any energies as well as the flux reduction. In figure b) the diffusion coefficient is scaled according to eq.~(3.2), 
 i.e., at higher rigidities the diffusion process is less efficient and, then, GCR spectrum is less modulated.}
 \label{fig:Comparison}
 \end{center}
\end{figure}

The numerical method presented in previous section was applied to GCR propagations into an ideal  spherical heliosphere of 100 AU. The LIS from 0.4 GeV to 100 GeV is taken from Ref.~\cite{Bobik2011ApJ}.
We test the solution with both approach assuming two different diffusion coefficients:
%
\begin{eqnarray}
\textrm{a)}&&  \kappa_{diff}=k_0; \label{eq_k0_a}\\
\textrm{b)}&&  \kappa_{diff}=k_0\beta  P;\label{eq_k0_b}
\end{eqnarray}
where $k_0\simeq 2.22 \times 10^{-4}$ AU$^2$ s$^{-1}$ is the diffusion parameter, $\beta$ is the particle speed in unit of light speed $c$ and $P=\frac{pc}{Ze}$ is the particle rigidity. 
In Fig.~\ref{fig:Comparison} we show the comparison of two solutions for the Forward-in-time and Backward-in-time case. For the Backward solution we used 1,000 quasi particles for each energy bin (70,000 total events generated), while for the forward case we simulate a total of $5\times10^{10}$ quasi-particles, that allows a simulated modulated spectrum with a statistical error lower than 1\%. The relative difference between the two solutions has a root mean square of $\approx 1 \%$, evaluated in the range 0.4-20 GeV. We modified $k_0$ from $1\times 10^{-4}$ up to $3 \times 10^{-4}$ AU$^2$ s$^{-1}$ with similar results, but with root mean square increasing in the worst case up to $5 \%$, leading to the conclusion that the systematic uncertain related to the Monte Carlo solution can estimated as less than $\sim$5\% in the considered range (a detailed study is ongoing).  

The descriptions of the algorithms clearly suggest that the \textit{backward-in-time} method is more efficient for evaluating the solution at one (or a small number of) point(s)~\cite{HandBookMC}. 

On other hand, if one is interested on GCRs spatial distribution, the Forward-in-time easily allows to evaluate multiple solutions inside the heliosphere domain with minor change in the algorithm. An Example of this study with HelMod can be found in Refs.~\cite{DellaTorre2013AdvAstro,ICRC13_DellaTorre}.

\section{Conclusion}
HelMod Model allows user to evaluate the modulated spectrum of GCR inside heliosphere solving the Parker's equation using a Monte Carlo Technique involving SDE. 
We developed two different algorithms solving the SDEs \textit{forward-in-time} and \textit{backward-in-time}. We show that the two solutions are equivalent and suitable to provide solutions of the modulated spectrum with a relative systematic uncertainty below of $5\%$, that is the typical error bar of GeV experimental data.

\acknowledgments
This work is supported by Agenzia Spaziale Italiana under contract ASI-INFN I/002/13/0, Progetto AMS-Missione scientifica ed analisi dati.
The PB and MP acknowledge VEGA grant agency project 2/0076/13 for support.

\providecommand{\href}[2]{#2}\begingroup\raggedright\endgroup
\end{document}